\documentclass{aa}
\usepackage{txfonts}
\usepackage{graphicx}
\begin{document}
\title{X-ray outburst of 4U 0115+634 and ROTSE Observations of its Optical 
Counterpart V635 Cas}
\titlerunning{4U 0115+634 / V635 Cas}
\authorrunning{Baykal et al.}
\author{Altan Baykal\inst{1}\thanks{\email{altan@astroa.physics.metu.edu.tr}} 
\and {\"U}mit K{\i}z{\i}lo\v{g}lu\inst{1} \and Nilg{\"u}n 
K{\i}z{\i}lo\v{g}lu\inst{1} \and \c{S}{\"o}len Balman\inst{1} \and S{\i}tk{\i} 
\c{C}a\v{g}da\c{s} \.{I}nam\inst{2}\thanks{\email{inam@baskent.edu.tr}}}
\offprints{A. Baykal}
\institute{Physics Department, Middle East Technical University,
  Ankara 06531, Turkey \and Department of Electrical and
Electronics Engineering, Ba\c{s}kent University, 06530 Ankara, Turkey} 
\date{Received / Accepted} 
\abstract{ROTSE IIId (The Robotic Optical Transient Experiment) observations of 
X-ray binary system 4U 0115+634/V635 Cas 
obtained during 2004 June and 2005 January make possible, for the 
first time, to study the correlation between optical and type II X-ray
outbursts. The X-ray outburst sharply enhanced after periastron passage where 
the optical brightness was reduced by 0.3 magnitude for a few days. We
interpret the sharp reduction of optical brightness as a sign of mass 
ejection from the outer parts of the disc of the Be star. 
After this sharp decrease, the optical brightness healed and reached the 
pre X-ray 
outburst level. Afterwards, gradual decrease of the optical brightness 
followed a minimum  then a gradual increase started again. Qualitatively, 
change of optical lightcurve suggests a precession of the Be star disc
 around a few hundred days. 
We also investigate the periodic signatures from the
archival RXTE-ASM (Rossi X-ray Timing Explorer - All Sky Monitor) light curve 
covering a time span of $\sim 9$ years.  
We find significant orbital modulation in the ASM light curve during the 
type I X-ray outburst.

}
\maketitle
\section{Introduction}
A "Be" star is an early type star which is close to the main sequence.
  Be stars
show Balmer emission lines and strong infrared excess in their spectra
(Slettebak 1988). Unlike normal B type stars, these properties suggest
that circumstellar material can form in a disc structure
around the Be star (Okazaki and Negueruela 2001). Indeed X-ray radiation in
some Be/X-ray binaries arises as a result of accretion of plasma from Be 
star to the compact object 
(Bildsten et al., 1997, Negueruela 1998,  Baykal et al. 2002).
 Fast rotation of the Be star, non-radial pulsations, and
 magnetic loops have been suggested as the causes that give rise to the disc
 around Be star, but it is not clear that any of them can explain the 
 observed transient nature of Be/X-ray binaries. Hanuschik (1996) suggested 
 that discs around Be stars are rotationally dominated and their motions 
are
quasi-Keplerian, however mechanisms for outflow from Be stars are
needed to explain the X-ray emission from Be/X-ray binaries
(Waters et al., 1988).

Some of the Be/X-ray binaries are persistent and relatively
low luminosity X-ray sources ($L_{x} \sim 10^{34}$ erg sec$^{-1}$). Their
X-ray luminosities vary up to a factor of 10. Most of the
Be/X-ray binaries show sudden (more than a factor of 10) increase
in their X-ray luminosities and are called Be/X-ray transients.

Be/X-ray transients show correlation between their orbital
periods and spin periods (Corbet 1986, Waters and van Kerkwijk 1989)
and exhibit two different kinds of outburst:

i) X-ray outbursts of low luminosity transients 
($L_{x} \sim 10^{36} - 10^{37}$ erg sec$^{-1}$) generally occur 
close to the periastron passages (type I X-ray outburst). The duration 
of these outbursts in most cases is related
to the orbital period (Okazaki \& Negueruela  2001).
ii)  On the other hand type II X-ray (or Giant) outbursts
($L_{x} > 10 ^{37}$ erg sec$^{-1}$) last several weeks or even months.
In most cases type II X-ray outbursts start after the periastron passage
but do not show any other correlation with orbital parameters
(Finger and Prince 1997).

The X-ray transient 4U 0115+634 (X0115+634) is an extensively
studied Be/X-ray binary system (Campana 1996; Negueruela et al. 1997;
Negueruela \& Okazaki 2001; Negueruela et al. 2001). The source was reported 
during the Uhuru satellite survey
(Giacconi et al. 1972; Forman et al. 1978). It was
also observed by  Vela 5B data base (Whitlock et al., 1989).
Using the observations of  SAS 3, Ariel V and HEAO-1 satellites,
the precise position of the X-ray source was determined
by Cominsky et al., (1978) and Johnston et al., (1978).
This location is used to identify the strongly reddened Be
star with a visual magnitude V $\sim $ 15.5 (Johns et al., 1978,
Hutchings and Crampton 1981) which was subsequently named as V635 Cas
(Khopolov et al., 1981). Using the SAS 3 timing observations,
the pulse period
(P$_{pul}$=3.6 seconds), orbital period (P$_{orb}$=24.3 days)
and the eccentricity (e=0.34) were found (Rappaport et al., (1978), see also 
Tamura et al., 1992).
\begin{figure*}[ht]
\resizebox{\hsize}{!}{\includegraphics[width=12cm,angle=-90]{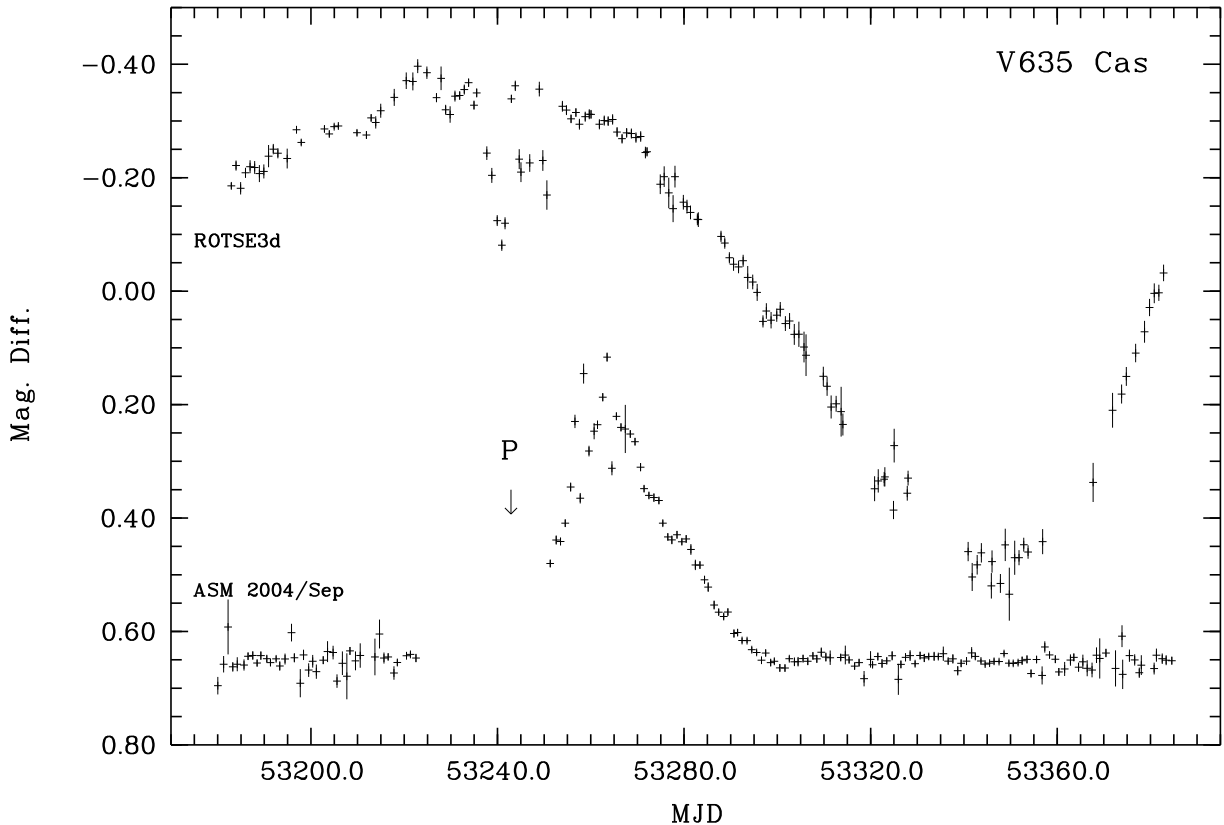}}
\caption{ROTSE and RXTE/ASM observations of V635 Cas and 4U 0115+634. Inset
arrow shows 
the periastron passage. X-Ray observations is plotted in arbitrary count 
rate units.}
\label{f1}
\end{figure*}

Multiwavelength long term monitoring observations of the optical counterpart
V635 Cas have shown that the emission lines and photometric magnitudes of
the Be star undergo quasi cyclic activity around $\sim$ 3-5 years
(Negueruela et al., 2001). This kind of activity suggests that
losing and reforming
a circumstellar disc around the Be star are possible.
 After each disc loss episode, the disc starts to reform and expands
 until it becomes unstable. The warping disk tilts and starts precessing.

 In this work we present new ROTSE observations of V635 Cas 
 (see also 
K{\i}z{\i}lo\v{g}lu et al. 2005 for preliminary results)
 and compare our results with
 RXTE/ASM observations of 4U 0115+634. 
 The comparison of light curves of the optical and
X-ray data have shown that type II outbursts in X-rays
enhanced significantly after the periastron passage.
The sharp decrease of the optical brightness
is interpreted as a sign of mass transfer episode to the compact object. 
The disk around the neutron star is formed in a few days after the
periastron passage. This is the first clear evidence of a 
correlation between optical brightness and X-ray outburst.   

\section{Observations}

\subsection{ROTSE}

The Robotic Optical Transient Experiment (ROTSE-III)
consists of four 0.45m worldwide robotic,
automated telescopes situated at different locations on Earth. 
They are designed for fast ($\sim$ 6 sec)
responses to Gamma-Ray Burst (GRB) triggers from satellites such
as Swift. Each ROTSE telescope has  a 1.85 degree of view imaged onto a 
Marconi
2048 $\times $ 2048 back-illuminated thinned CCD. These telescopes 
operate without filters, and have wide passband which peaks 
around 550 nm (Akerlof et al. 2003). ROTSE III telescopes are 
scheduled to observe optical transients when there
are no GRB events. In this work, we present optical observations of V635
performed by ROTSE IIId telescope
located at Turkish National Observatory (TUG) site, Bak{\i}rl{\i}tepe, 
Turkey. The observations took place
between MJD 53180 (June 2004) and MJD 53383 (January 2005).

A total of about 1850 CCD frames were analyzed.
After finding the instrumental magnitudes (Bertin$\&$Arnouts, 1996)
  ROTSE magnitudes were calculated by comparing all the
 field stars to the USNO A2.0 R-band catalog.
All the processes were done in sequential automated mode.
Barycentric corrections were made to the times of each observation
by using JPL DE200 ephemerides.

Fig. 1 shows the daily averages of data for V635 Cas
obtained with ROTSE IIId telescope.
The difference in ROTSE magnitudes of V635 Cas and the comparison star
( RA= 01$^{h}$ 17$^{m}$ 35$^{s}$.7, $\delta$= +63$^{\circ}$ 41$'$
44$''$) were plotted. As a check star, we used the one with
RA= 01$^{h}$ 18$^{m}$ 31$^{s}$.3, $\delta$= +63$^{\circ}$ 47$'$
30$''$.4.
\subsection{RXTE/ASM}

 The All Sky Monitor (ASM) on board of Rossi X-ray Timing Explorer (RXTE)
satellite consists of three wide-angle Scanning Shadow Cameras (SSCs).
These cameras are mounted on a rotating drive assembly, which cover
$\sim $70 $\%$ of the sky every 1.5 hours.
A detailed information of
the ASM can be found in Levine et al., (1996).
ASM data products can be found in the public archive in
 three different energy bands
(1.3-3.0,3.0-5.0,5.0-12.0 keV). In Fig. 1, we present the
ASM light curve of 4U 0115+634 in the 5.0-12.0 keV energy band 
together with the light curve of optical counterpart
V635 Cas.

\section{Results and Discussion}

In this work, we present RXTE/ASM observations of 4U 0115+634
and its optical counterpart V635 Cas. A general view of the evolution of 
the
unfiltered optical and RXTE/ASM light curve is shown in Fig. 1.

Negueruela et al (2001) showed that the general optical, infrared and
X-ray behaviour of this system can be explained by the dynamical
evolution of the viscous circumstellar disc around V635 Cas.
The evolution of emission lines and photometric magnitudes
indicated that losing and reforming a circumstellar disc around the
V635 Cas are quite possible in timescales of
($\sim3-5$) years. After each disc-loss episode, the disc starts
to reform in less than 6 months.
The disc surrounding
the V635 Cas is truncated at a resonance radius depending on the viscosity
parameter by the tidal/resonant interaction with the neutron star
(Negueruela  and Okazaki 2001, Negueruela et al 2001). Then
the disc becomes unstable to warping, tilts and starts precessing
(Porter 1998).

Typical timescale of optical variations seen in Fig. 1 is much less
then a few years. Therefore we  consider the variation 
in the optical brigthening as being closely related to the 
precession of circumstellar disk.
However some of the short time scale variations could be
related to the dynamical
instabilities of the viscous circumstellar disc around V635 Cas.
Indeed, the interaction of the disc with the neutron star causes
small variations in the light curve which is clearly seen in our observations.
The optical brightness sharply decreased by  $\sim$ 0.3 magnitude during
this episode. After this sharp decrease,  the optical brightness healed with
small variations on a time scale of several days. It is interesting to 
note
that the optical brightness resumed its pre-instability value.

If the decrease in the optical brightness associates with the mass loss
episode to compact object then the change in optical
magnitude ($\Delta m \sim 0.3$)  corresponds
to $\dot M \sim 10^{-8} M_{\odot} yr^{-1}$ mass loss rate
from the V635 Cas.
This material can be captured and form an accretion disk as suggested by
Negueruela et al., (2001) to produce the observed
type II X-ray outburst with a mass accretion rate of
 $\dot M_{x} \sim 4 \times 10^{-9} M_{\odot} yr^{-1}$ (Coburn et al., 
2004).

Therefore it is quite resonable to
explain the observed light curve as a sign of the precession
of the disc around the Be star. Due to the precession of the warped disc, 
projection of the disc on to the plane of the sky changes giving rise to the change 
of light coming from the system.  When the denser, elongated part of the disc
is in the line of sight, light output coming from the system  is less.
Close to the periastron passages, the outer edge of the precessed disc may 
cause type II X-ray outbursts. Otherwise the disc continues to
precess without causing any type II X-ray outburst.

\begin{figure} [t]
\resizebox{\hsize}{!}{\includegraphics[angle=-90]{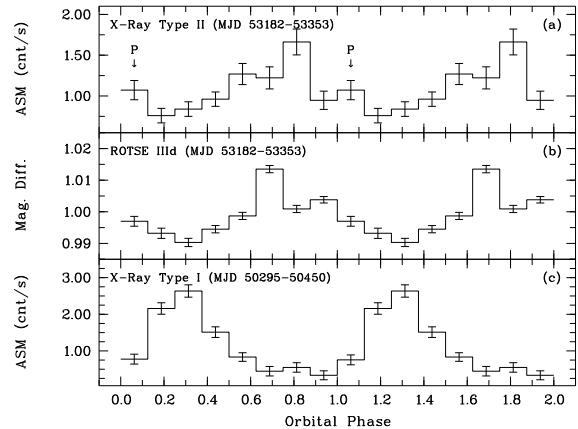}}
\caption{ Folded RXTE/ASM/X-ray  lightcurve corresponding to
time span of the optical outburst
($\sim$ MJD 53182 - 53353) and folded ROTSE IIId lightcurve covering the same
time interval are presented at the upper {\bf{(a)}}
 and middle panels {\bf{(b)}} respectively.
  Bottom panel {\bf{(c)}} presents the folded light curve during the type I
  RXTE/ASM/X-ray outburst   (MJD 50295-50450).
 Phases 0.0625 and 1.0625 are the periastron passages
  denoted by arrow P. }
\label{f2}
\end{figure}

Negueruela et al. (1998) investigated the ASM light curve between
MJD 50087-50530. In their analysis, they found that there
was type I X-ray outburst activity between MJD 50295-50450 and folded 
profile showed that type I X-ray outburst peak  $\phi \sim 0.3$ away 
from the periastron. In order to see orbital period signature of 
a type II X-ray outburst, we folded the optical outburst
light curve (MJD 53182-53353) and the 
corresponding ASM observations on the orbital period of 24.31 days. As seen
from Fig. 2a,b, both folded lightcurves agree with each other.
It is interesting to note that
even if the outburst in X-rays starts a few days after the periastron,
peak values of both profiles occur at $\phi \sim 0.25$ ($\sim 6 $ days)
before the periastron passage. This suggests that
accretion disk enhances $\sim 6 $ days
prior to the periastron passages. Further observations are needed to 
confirm this behaviour.
\begin{figure*}
\includegraphics[width=12cm, angle=-90]{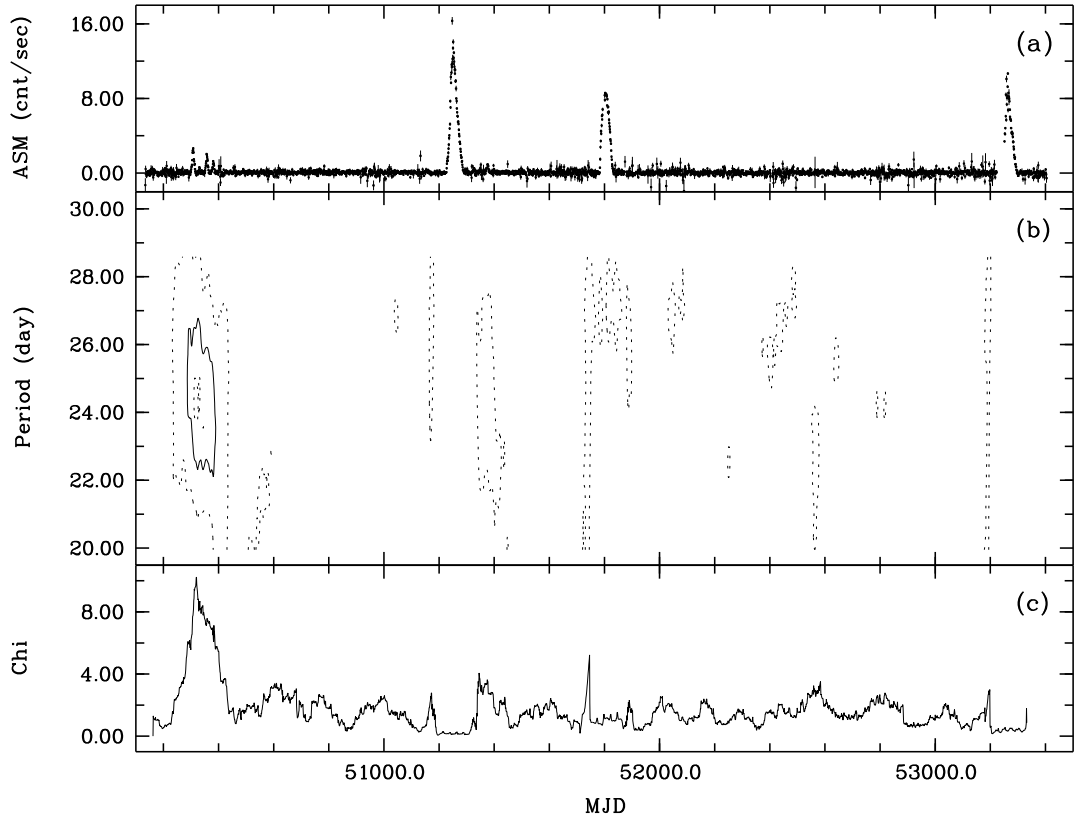}
\caption{{\bf{(a)}} ASM light curve of 4U 0115+634.
 {\bf{(b)}} Dynamical orbital period search of ASM light curve.
 Contour lines denote 1, 3 and 4$\sigma$ confidence levels.
 Dashed lines denote 1$\sigma$, solid line 3$\sigma$ and the contour
 inside  4$\sigma$.  (see the text for details).
 {\bf{(c)}} Reduced $\chi ^{2}$ values for folding at orbital period 24.31 days.
 (note that 8 phase bins are used for folding) are shown in the lower
 panel.}
\label{f3}
\end{figure*}

In order to see the statistical significance of these profiles  and
search for other possible orbital signatures in the ASM light curve,
 we apply $\chi ^{2}$
orbital period search. In this procedure we compute dynamic
period search by calculating $\chi ^{2}$ values from 150 days ($\sim$ 6 
orbital cycle)
interval with a new interval beginning every 1 day.
The resulting orbital search is  not independent since the data segments
overlap, but this method identifies the range of times for which the 24.31 days
orbital period is present. We search orbital
periods between 20 and 30 days and present their $\chi ^{2}$ distributions
at the mid value of time intervals.
In Fig. 3, we present the detection
contours for $\Delta \chi ^{2}$ statistics. As seen in Fig. 3,
statistically significant orbital period modulations are only present
around the type I X-ray outburst region as suggested by Negueruela et al. (1998).
Eventhough orbital period signatures around
type II X-ray outburst regions are not significant (around  $\sim 1\sigma $
confidence level), it is important to note that, as seen in Fig. 2,
both optical and X-ray profiles agree with each other 
during the type II X-ray outburst (MJD 53182-53353).
On the other hand profiles of type I X-ray and type II X-ray
 outbursts are phase shifted 
by $\sim 198$ degrees ($\Delta \phi \sim 0.55$) (see fig. 2 a,c).
 This phase shift and weak orbital
signature strongly suggest that the accretion mechanisms of the two types of
outburst are geometrically different.
In type I X-ray outbursts accretion is moderate while in
type II X-ray outbursts extended
nature of the Be disc and the high accretion rates
are not affected significantly by the eccentric orbit
of the binary system. Therefore it is quite natural to see
stronger orbital modulation in type I X-ray outbursts relative to the
type II X-ray outbursts.

 In ROTSE observations,
  the change in the optical brightness
 is related to the dynamical evolution of the viscous decretion disc
 surrounding the Be star V635 Cas.
 Some of the disc loss and reformation cycles cause type II X-ray
 outburst. During the disc growth phase, decretion disc becomes unstable 
 and the radiation driven warping starts (Porter 1998). As the disc warps,
 the tilt of the outer regions increases and the disc starts to precess.
 The precession period for this source is suggested to be of the order of
  100 days (Negueruela et al., 2001).

Future monitoring of V635 Cas with the ROTSE IIId telescope may yield 
better understanding of the precession and the outburst behaviour of 
this source.

\begin{acknowledgements}
We thank international ROTSE collaboration and Turkish National Observatory for
the optical facilities. We acknowledge RXTE-ASM team for the X-ray
monitoring data.
\end{acknowledgements}


\begin{thebibliography}{}
\bibitem{} Akerlof, C.W., Kehoe, R/L/. McKay, J.A., 
           Rykoff, E.S., et al., 2003, \pasp, 115, 132
\bibitem{} Baykal, A., Stark, M., Swank, J., 2002, \apj, 569, 903 
\bibitem{} Bertin, E., Arnouts, S., 1996, \aaps, 117, 393
\bibitem{} Bildsten, L., Chakrabarty, D., Chiu, J., et al., 
           1997, \apjs, 113, 367
\bibitem{} Campana, S., 1996, \apjs, 239, 113
\bibitem{} Coburn, W., Kalemci, E., Kretschmar, P., et al., 2004, Atel, 
           337, 1C  
\bibitem{} Cominsky, L., Clark, G.W., Li, F. et al. 1978, \nat, 273, 367
\bibitem{} Corbet, R.H.D., 1986, \mnras, 220, 1047
\bibitem{} Forman, W., Jones, C., Cominsky, L., et al., 
           1978, \apjs, 38, 357
\bibitem{} Finger, M.H., Prince, T.A., 1997, Proc. Fourth 
           Compton Symp. 1(AIP, Woodbury, NY), 57 
\bibitem{} Giacconi, R., Murray, S., Gurksky, H., et al., 
           1972, \apj, 178, 281 
\bibitem{} Hanuschik, R.W., 1996, \aap, 308, 170
\bibitem{} Hutchings, J.B., Crampton, D., 1981, \apj, 247, 222
\bibitem{} Johns, M., Koski, A., Canizares, C., et al., 1978, 
           IAUC, 3171 
\bibitem{} Johnston, M., Bradt, H., Doxsey, R., et al., 
           1978, \apj, 223, L71
\bibitem{} Khopolov, P.N., Samus, N.N., Kukarkina, N.P., 
           Medveddeva, G.I., Perova, N.B., 1981, IBVS, 2042
\bibitem{} K{\i}z{\i}lo\v{g}lu, {\"U}., Baykal, A., 
           K{\i}z{\i}lo\v{g}lu, N, 2005, IBVS, 5590   
\bibitem{} Levine, A.M., Bradt, H., Cui, W., et al., 1996, \apj, 
           469, L33  
\bibitem{} Negueruela, I., 1998, \aap, 338, 505
\bibitem{} Negueruela, I., Grove, E.J., Coe, M.J., et al., 1997, 
           \mnras, 284, 859 
\bibitem{} Negueruela, I., Reig P., Coe, M. J., Fabregat, j.,
          1998, \aap, 336, 251
\bibitem{} Negueruela, I., Okazaki, A.T., 2001, \aap, 369, 108
\bibitem{} Negueruela, I., Okazaki, A.T., Fabregat, J., 
           Coe, M.J., Munari, U., Tomov, T., 2001, \aap, 369, 117    
\bibitem{} Okazaki, A.T., Negueruela, I., 2001, \aap, 377, 161
\bibitem{} Porter, J.M., 1998, \aap, 348, 512
\bibitem{} Rappaport, S., Clark, G.W., Cominsky, L., 
           Joss, P.C., Li, F., 1978, \apj, 224, L1   
\bibitem{} Slettebak, A., 1988, \pasp, 100, 770
\bibitem{} Tamura, K., Tsunemi, H., Kitamoto, S., et al., 
           1992, \apj, 389, 676
\bibitem{} Waters, L.B.F.M., van den Heuvel E.P.J., Taylor, A.R.,
           et al., 1988, \aap, 198, 200
\bibitem{} Waters, L.B.F.M., van Kerkwijk, M.H., 1989, \aap, 223, 196
\bibitem{} Whitlock, L., Roussel-Dupre, D., Priedhorsky, W., 1989, 
           \apj, 338, 381


\end{thebibliography}
\end{document}